\documentclass[manuscript,screen]{acmart}
\usepackage{caption}
\AtBeginDocument{%
  \providecommand\BibTeX{{%
    \normalfont B\kern-0.5em{\scshape i\kern-0.25em b}\kern-0.8em\TeX}}}

\copyrightyear{2026}
\acmYear{2026}
\setcopyright{cc}
\setcctype{by}
\acmConference[CHI EA '26]{Extended Abstracts of the 2026 CHI Conference on Human Factors in Computing Systems}{April 13--17, 2026}{Barcelona, Spain}
\acmBooktitle{Extended Abstracts of the 2026 CHI Conference on Human Factors in Computing Systems (CHI EA '26), April 13--17, 2026, Barcelona, Spain}
\acmDOI{10.1145/3772363.3798675}
\acmISBN{979-8-4007-2281-3/2026/04}

\begin{document}

\title{Not Just Duolingo: Supporting Immigrant Language Preservation Through Family-Based Play}

\author{Alejandro Ciuba}  
\email{alejandrociuba@pitt.edu}
\affiliation{%
  \institution{University of Pittsburgh}
  \city{Pittsburgh}
  \state{Pennsylvania}
  \country{USA}
}
\author{Zheng YY Li}
\email{ZHL195@pitt.edu}
\affiliation{%
  \institution{University of Pittsburgh}
  \city{Pittsburgh}
  \state{Pennsylvania}
  \country{USA}
}
\author{Aakash Gautam}
\email{AAKASH@pitt.edu}
\affiliation{%
  \institution{University of Pittsburgh}
  \city{Pittsburgh}
  \state{Pennsylvania}
  \country{USA}
}








\renewcommand{\shortauthors}{Ciuba, Li and Gautam}  

\begin{abstract}
For immigrants, language preservation is crucial to maintain their identity, but the process of immigration can put a strain on a community's ability to do so. We interviewed eight Nepali immigrants to understand barriers to language preservation across sociopolitical contexts in Nepal and immigrant life in the United States. Participants described strong motivation but limited institutional support, time and resource constraints, and English-dominant environments that widen parent-child language gaps. They envisioned technology that supports interactive, family centered learning. In response, we are developing an audio-first, point-and-click language learning game based on the theory of comprehensible input, designed for parent-child co-playing. An early evaluation with four design experts reveals promising gameplay, and the need to simplify symbol-heavy UI. We conclude with implications for designing language technologies that support preservation through relations while acknowledging the limits of design.  

\end{abstract}


\begin{teaserfigure}  
    \includegraphics[width=\linewidth]{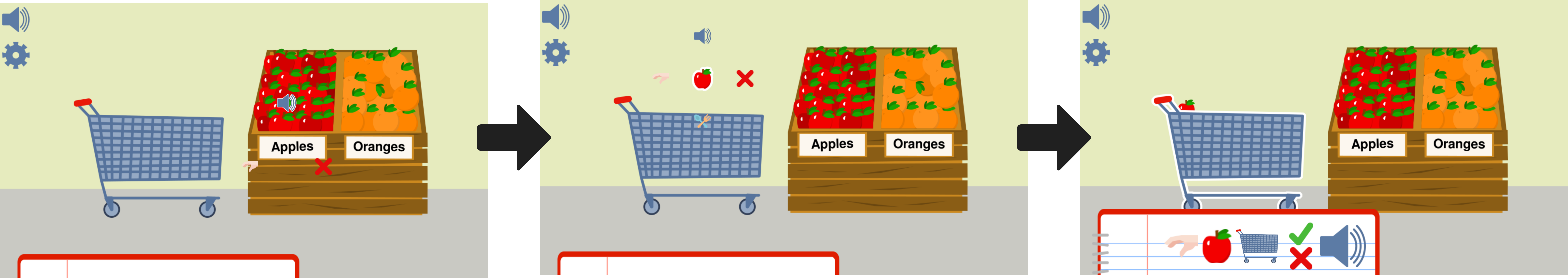}
\Description{The game prototype. There are three images displayed left to right. The top right corner of each image has a speaker button and a gear button. In the first image, there is a shopping cart and a produce shelf containing apples on the left side and oranges on the right side, all against a plain background with a beige floor and a tan wall. The second image is the same, but now an apple is hovering over the cart with four menu options displayed around it. From the right going clockwise, the menu has a red X symbol, a crossed fork and spoon, a hand, and a speaker at the top. The final image has the apple in the shopping cart with no menu. In the bottom left half, there is a large notebook. Displayed in the notebook is a hand, an apple, a shopping cart, a green check button with a red X below it, and lastly a speaker button.}
    \caption{Our audio-first game prototype to support co-presence in intergenerational language learning. Players explore the environment by (1) clicking objects and viewing available actions, (2) performing an action sequence, which is then (3) narrated in the action panel.}
    \label{fig:game}
\end{teaserfigure}


\maketitle

\section{Introduction}
\label{introduction}

Language preservation and promotion efforts vary drastically with respect to diverse sociopolitical factors \cite{hinton2018a, hornberger2023a}.
Amid a  worldwide decrease in linguistic diversity \cite{skutnabb-kangas2010, grenoble1998}, sustaining Indigenous heritage languages is a critical challenge.
In Nepal, the linguistic diversity, political prioritization of a single language, and uneven institutional support have limited formal curricula for Indigenous languages and thus pushed the onus of preservation on families \cite{phyak2021subverting, toba2010diversity}.
For Nepali immigrants in the US, and immigrant families more generally, this challenge is compounded by the fact that children quickly adapt to English-dominant schools and peer environments, while parents face pressures to assimilate in foreign spaces \cite{salari2025, ponizovsky2004}.
This risks weakening interpersonal connections as the parents' language of emotional expression diverges from that of their children \cite{han2023case, harris2023acculturation}.
Language, after all, is a medium for intimacy and care, nurturing intergenerational connections to family and heritage \cite{lopez-class2011, harris2023}.



Community-based efforts to preserve and promote immigrant languages in the US are limited by assimilation-oriented policies, unstable funding, and reliance on informal support networks \cite{rumbaut2013immigration, ricento2005problems}. 
In this space, immigrant families rely on digital language tools, which implicitly frame learning as an individual engagement through standardized and often decontextualized content \cite{wang2021dismantling, burston2015twenty}. 
For families seeking connections with each other and their heritage, current efforts are misaligned. 
To explore these issues, we interviewed eight Nepali immigrants in the US. 
The interviews highlighted the relational aspect of language acquisition, which led us to prioritize interactive experiences that facilitate family-centered opportunities to hear and use the language. 
We developed an early game-based design that focuses on immersive comprehensible input \cite{rodrigo2004, alatis1992} and parental guidance to foster language learning.

Our work contributes an account of the motivations and challenges Nepali immigrants face in preserving Indigenous languages.
From this, we build an early game prototype and draw implications for interactive systems that support intergenerational language preservation.
In particular, we emphasize the importance of designing for co-presence and prioritizing oral modalities when promoting language preservation. 

\section{Related Work}
\label{related-work}
Our work sits at the intersection of HCI research with immigration populations and game-based learning in families. 

\subsection{Culture and Language Preservation with Migrant Populations in HCI}
\label{culture-and-language-preservation-with-migrant-populations-in-hci}


Previous research in HCI and CSCW has examined technologies that support immigrants' immediate needs and navigation of unfamiliar institutions, and how immigration affects identity, belonging, and daily life (e.g., \cite{georgiou2023, pei2020, tachtler2020, chang2024, brown2012}).
Despite this, the development of cultural preservation technologies for and with immigrant communities remains comparatively understudied \cite{sabie2022, sabie2020}, with much of the work exploring memory, storytelling, and speculative approaches to technology-driven cultural preservation \cite{sabie2020, chowdhury2024, shen2025, nisi2025, bala2024}, with few notable exceptions \cite{lewis-yang-2012-building, ariyasinghe2025}.
Building on these, we see heritage language preservation as an inherently relational practice that must be embedded into familial and social contexts, and thus, we examine Indigenous language preservation in a family-centered, oral context.

\subsection{Games for Language Learning and Parent-Child Co-Engagement}
\label{game-design-for-cultural-and-language-education}

Games are a popular medium to share and celebrate cultures (e.g., \cite{shen2025, ariyasinghe2025, kalinda2025, baranoski2015, malegiannaki2020}) as their interactivity allows an audience connection that is unique to the medium \cite{chen2013}. 
Thus, the medium has been used for culturally-grounded learning (e.g., \cite{bermudez2023, cheng2017}), including language learning through audio-focused, immersive comprehensible input  that slightly challenges learners within their zone of proximal development \cite{cho2021, vygotsky1978mind, rodrigo2004}.
However, incorporating comprehensible input requires balancing gameplay complexity and educational load \cite{cho2021, nadolny2017, martin2014}.
Sociocultural learning and joint media engagement research also shows that parents' involvement in their child's play can support educational and creative goals if the parents' role is negotiated and well established \cite{druga2022, yu2023, einarsson2022}.
Additionally, activity focus, generational differences in technological literacy, and cultural practices all influence family dynamics when engaging in jointly experienced cultural learning activities \cite{jiang2025}.
Building on the scholarship, we consider comprehensible input and parent-child dynamics in our language learning game as potential sites for heritage language preservation. 

\section{Methodology}
\label{methodology}
%




We interviewed eight participants before finding theoretical saturation \cite{hennink2017code}.
All participants are Nepali immigrants who are members of at least one Nepali Indigenous community with their own language. They represented several Indigenous language communities --- Newari (P2, P4, P6), Maithili (P1, P7), Limbu (P3), Sherpa (P5), and Magar (P8).
They resided across different regions of the United States.
All participants spoke Nepali and English, and all spoke their Indigenous language(s), except for P2 and P8, who stated they understood but could not speak the language.

The study received IRB approval prior to data collection.
The interviews focused on the participants' experiences with language and cultural preservation in both Nepal and the US and were audio recorded, totaling around 7.5 hours.  
These recordings were transcribed and coded to produce 1457 codes \cite{braun2006using}. 
Through iterative comparison across transcripts, the first two authors, in discussion with the last author,  merged related codes into intermediate categories and refined them into higher-level themes as patterns stabilized.
\section{Findings and Design}
\label{findings-and-design}

We find that the participants' desire for language preservation arises from personal identity but is challenged by environmental contexts, and that their ideas for technological interventions focus on play and interactivity.




\subsection{Interview Findings: Why a Family-Involved Game}
\label{interview-findings}

\subsubsection{Assimilation Pressures and Intergenerational Gaps}
\label{assimilation-pressures-and-intergenerational-gaps}

In the United States, immigrants described pressures to assimilate that can suppress Indigenous language use.
P4 reflected, \textit{``Maybe it is an identity crisis, or survival mode . . . I cannot use the American mentality in Nepali society and vice-versa.''}
P3 shared similar thoughts, ``\textit{When I tried to apply cultures I learned in Nepal [in the U.S.], I could not assimilate. 
Then I became someone who wants to be like Americans, learn their culture.}''
They linked these pressures to parenting practices and language choices.
For example, P8 shared that immigrant parents may prioritize English fluency because \textit{``[parents] want their children to speak like native English speakers . . . [they] realize their struggle in a new place because of English.''} 

However, they also acknowledged that language loss weakens intergenerational relationships. 
For instance, P6 shared, \textit{``The other reason I want them to know Newari is for them to be able to talk to their grandparents. I don't want them to talk in English.''}
P8 described the emotional gap that arises from language differences, \textit{`` . . . when immigrant parents speak English with their children born in the USA, there is a gap when it comes to expressing the emotion from the parents' end because their comfort language is different from English . . . ''}
We note that these experiences highlight language preservation as a means of sustaining family connections under immigration pressures. 

\subsubsection{Language Learning Pushed Into the Home}
\label{language-learning-pushed-into-the-home}

Participants described learning language largely outside of formal schooling due to political and infrastructural issues. 
Participants acknowledged the political challenges in reforming Indigenous languages in Nepali formal education. 
As P3 explained, \textit{``. . . I have my doubts because Nepal, although small, is a diverse country . . . adding in curriculum might raise questions from people why their language is not being represented.''} 

P2 noted that even when their language is spoken, it is often \textit{``within the family, inside that house only.''} 
There was an emphasis on oral tradition to acquire language as P4 shared, \textit{``The majority of Limbu learned the language by listening, no one has learned it in an academic environment because of the political scenario, scripts, language were not taught.''}
P3, who spoke Limbu language, corroborated, \textit{``I was not taught the language in school either . . .  I used to hear grandmothers speak it since when I was a child, and I learned the language based on just listening.''}
Without equitable formal curricular resources and limited institutional support, families are the primary site for Indigenous language preservation.

\subsubsection{Motivations for Language Learning} 
\label{motivations-for-language-learning}

Participants emphasized that the responsibilities of daily life can demotivate individuals from pursuing language preservation.
P8 shared that while some people are involved in preservation efforts, for others, language learning materials are \textit{``[an] extra expense and demanding of their time'', adding that it requires significant ``personal commitment.''}
P7 shared, \textit{``I don't think I would be convinced if someone tried to put me in a language class at this age . . . ''}, emphasizing that their lived realities created a situation where \textit{``the importance and respect that needs to be held for languages fades away . . . ''}
We see this as a justification for involving children and families. 

At the same time, participants described connection to elders and family history as a powerful motivator. 
P3, for example, shared, \textit{``I feel amazing talking to my seniors using my language, but I feel upset as well that I cannot speak it fluently.''}
P1 shared their pride in learning a language that was intertwined with cultural history, \textit{``My grandfather was actively involved in this language preservation . . . I would find his notepad where the history of Maithili language . . . were written. 
I learned about the history of the language from there.''}
This relational element motivates language learning but raises a design tension for immigrant communities, whose immigration often limits opportunities to foster broad familial relationships and connections to heritage \cite{sabie2022}. 


\subsubsection{Technology Interventions for Language Preservation}
\label{technology-interventions-for-language-preservation}
When participants envisioned possibilities with technology, they emphasized interactive and family-centered experiences. 
P4 suggested creating a rich context because, \textit{``Children also learn from hearing those languages in their surroundings. If there could be an environment for kids to be in those surroundings, it could be playing games''}, adding further that \textit{``parents should also be involved in it.''}
Similarly, P6 suggested using oral, story-based materials, \textit{``Rather than focusing on writing scripts, maybe we can start sharing the stories out of there.''}
P5 echoed this sentiment when ideating on technological interventions, stating, \textit{`` . . . there would be a native speaker to teach users how to pronounce.''}

Participants also cautioned against replicating existing language apps that focused on machine-centric, individual learning. 
As P4 put it, \textit{``I do not think we should build something that’s Duolingo from 2012, but there could be something where people can live and interact, chat with others rather than machines.''}
This emphasis on parent-inclusive, interactive learning motivates our design work that supports listening and parent-child co-participation.

\subsection{Design Intervention: Audio-First, Low-Barrier, Parent-Child Play}
\label{design-intervention}

\subsubsection{Prototype Overview and Core Design Choices}
\label{game-design}

Based on our findings above, we developed an audio-first, point-and-click language learning game (see Figure \ref{fig:game}) intended to support parent-child co-present learning. 
The game draws on the theory of comprehensible input \cite{tarekhamza2016, rodrigo2004} to emphasize listening comprehension. 
To reduce game complexity and support users with diverse digital skills, we use point-and-click mechanics \cite{inkpen2001}. 
We envision the child as the primary player, with the parent serving as a mentor who scaffolds learning while bonding through shared play, aligning with the participants' emphasis on family as a primary site of learning. 

We implemented the prototype with the Godot Engine \cite{linietsky}. 
The game includes translations for Spanish and Mandarin Chinese to test the game mechanics; future work will involve Nepali Indigenous languages.
We use a grocery-shopping setting, which is a familiar scenario, to prioritize interaction design before incorporating culturally specific story content, which we plan to focus on once the game is brought to the community.
\looseness -1

When the game begins, the player listens to an audio prompt in the target language describing a task (e.g., ``you put an apple in the shopping cart'').
Interactable objects glow when the player's mouse hovers over them.
Clicking an object reveals a radial menu with its available actions.  
Players can elect to hear an object's audio label and select from the object's list of item-specific actions (e.g., eating an apple).  
When the player completes an action, the relevant icons appear in the "action panel", which forms a sentence.
Players can then listen to their constructed sentence, submit it, or clear the action panel. 
Feedback is provided through ``correct'' or ``incorrect'' jingles and then transitions to the next scene. 
\looseness -1


\subsubsection{Early Prototype Evaluation}
\label{early-prototype-evaluation}

We reached out to three HCI researchers (one with educational design experience and two with game design experience) and a game designer for playtesting to identify major usability issues before running parent-child studies.
The sessions lasted 30 to 60 minutes. 
Each tester was briefed on the game's intended audience and the mechanics.
They were encouraged to play the game in a language they did not speak (three selected Spanish and one selected Chinese).
We intentionally did not explain the action symbols to evaluate if the interface was learnable.
After they played, we used the EOTA method \cite{hammer2018eota} to elicit structured feedback, which we combined with our field notes to conduct a round of thematic analysis. 


The testers enjoyed playing the game and successfully used the object audio to infer the task audio.
However, two testers reported that later levels became too easy due to pattern repetition. 
Some noted confusion with the radial menu symbols (e.g., ``eat'' using a utensil symbol and how to pick up objects).
Critically, all testers found the action panel difficult to interpret due to its heavy reliance on symbols. 
The testers suggested having an object's audio play automatically when clicked, reducing information load in the action panel and refining the symbols.
Importantly, when reflecting on the intended audience, the testers recommended deepening parent engagement through participation rather than observation. 
This feedback aligns with our findings and shows how gaps in our design can be addressed to better meet our participants' ideals.
\looseness -1


\section{Discussion}
\label{discussion}

We find that Nepali immigrants center family connection and cultural identity in language preservation efforts, and that compounding pressures make language preservation difficult. 
Our game prototype responds to these findings by centering parent-child engagement.
Here, we reflect on two implications for design. 
\looseness -1
%



\subsection{Designing for Relational Language Preservation} 
\label{design-implications}

Participants consistently described language preservation as a practice embedded in family relationships and did not see language as a skill to acquire, but rather as a medium to build intergenerational connections. 
For example, when P8 explained the gap in expressing emotions arising from language differences (Section \ref{assimilation-pressures-and-intergenerational-gaps}), we read it as a rupture in relationship. 
This relational framing has implications for how we design language preservation technologies. 

\subsubsection{Designing for Co-Presence:} 
We find that mainstream language learning technologies prioritize individualized language acquisition \cite{burston2015twenty, wang2021dismantling}.
Our participants instead advocated for interactive language learning experiences where families can engage together, encapsulated by P4's statement not to design another Duolingo (Section \ref{technology-interventions-for-language-preservation}).
The unit of design should be the dyad or the family rather than the individual learner.
In design, this reframes metrics of success to deprioritize rote memorization or time on task in favor of interaction quality between family members and the extent to which the play fosters conversations in the target language. 
Importantly, co-presence should not be assumed; it must be scaffolded through design. 
This insight emerged during our prototype evaluation when testers noted that our initial design restricted parental involvement. Our future iteration will offer parents a clearer role as guides and co-players while also including mechanisms to prevent them from overtaking the child's agency \cite{einarsson2022}. 
From here, we see the challenge as creating interdependence where neither the child nor the parent can succeed alone.
\subsubsection{Prioritizing Oral Modalities:}
Many Indigenous languages have oral traditions (e.g., \cite{mahuika2019rethinking, leonard2007deg, setiawan2024role}).  
Our participants described learning through listening, such as by hearing their grandmothers speak (Section \ref{language-learning-pushed-into-the-home}).
An audio-first approach honors oral traditions and histories while reducing orthographic barriers for communities, promoting equity. 
We follow the principle of ``low-floor, high-ceiling'' \cite{papert2020mindstorms} in our design.
To lower the floor, we designed simple point-and-click mechanics that minimize prerequisite skills. 
Players can listen, click, and construct meaning from familiar contexts. 
We believe this supports users with varying digital literacies, such as parents who may be less comfortable with technology, while focusing on language rather than game mechanics.
The high ceiling in our design emerges from relational depth where, as families develop fluency together, the game becomes a site for storytelling and intergenerational knowledge transmission 
beyond the prototype's current tasks. 

\subsection{Limitations and Future Work}
\label{limitations-and-future-work}

\subsubsection{The Limits of Game Design:}
\label{the-limits-of-game-design}
Nepal's linguistic politics have limited formal curricula for Indigenous languages \cite{phyak2021subverting, toba2010diversity}, with immigration further severing community networks, and making resources for Indigenous language learning scarce. 
Families thus bear the burden of preservation. 
We acknowledge the limits of design in this context. 
Our game is intentionally designed for low-effort engagement through short sessions, directly addressing participants' limited ability to dedicate time and resources.
It is not a substitution for the institutional and infrastructural support that the participants lack, but rather we see it as one viable approach to make some impact, given the constraints.
Indeed, we caution that design interventions which place responsibilities on individuals and families risk further reinforcing the lack of institutional responsibility they are intended to address. 
Therefore, we urge other HCI researchers in this space to examine how their designs distribute the labor of preservation and to resist framings that romanticize individual or family resilience while ignoring structural factors.

\subsubsection{The Limits of Our Methodology:}
\label{the-limits-of-our-methodology}
Our participants were all adults; we did not have children's perspectives.
Our findings are specific to Nepali immigrants from Indigenous communities and may not generalize to other populations.
Our prototype was evaluated by design experts, which is a necessary step before community trials, but is a limitation nonetheless. 
Future work will address current feedback, record Indigenous language audio with native speakers, develop culturally grounded content through community collaboration, and conduct participatory design with immigrant families. 
\looseness -1

\section{Conclusion}
\label{conclusion}



We examined how Nepali immigrants from Indigenous communities described language preservation amid assimilation pressures, limited institutional support, and the practical demands of everyday life. 
Participants framed language as a relational practice tied to intergenerational connections, positioning the home as the primary site for language learning. 
In response, we designed an audio-first, family-centered language learning game that supports parent-child play. 
While our prototype is at an early stage, it responds to participants' call for alternatives beyond standardized language apps by emphasizing family connection as a mechanism for sustaining Indigenous languages.
\looseness -1

\begin{acks}
    We would like to thank the participants for their time and insights.
    This research was partly supported by Google’s Award for Inclusion Research, awarded to Aakash in 2022.
\end{acks}

\bibliographystyle{ACM-Reference-Format}
\bibliography{references/alejandro-references, references/references}


\end{document}